\documentclass[12pt,a4paper]{article}
\usepackage[german, english]{babel}
\usepackage{a4wide}
\usepackage{amsmath}
\usepackage{amssymb}
\usepackage{ifthen}
\usepackage{epsfig}

\begin{document}
\begin{titlepage}
\title{Collective coordinates of the Skyrme model coupled with fermions}

\author{Ya.~Shnir \\ [4mm]
{\it {\footnotesize 
Institut f\"ur Physik, Universit\"at Oldenburg}}\\
{\it {\footnotesize
D-26111, Oldenburg, Germany
}}
\date{~}
}
\maketitle 
\begin{abstract}
The problem of construction of fiber bundle over the moduli
space of the Skyrme model is considered. We analyse an 
extension of the original Skyrme model which includes  
the minimal interaction with fermions. An analogy with modili space of the 
fermion-monopole system is used to 
construct a fiber bundle structure over the skyrmion moduli space. 
The possibility of the  
non-trivial holonomy appearance is considered. It is shown that the effect of
the fermion interaction turns the $n$-skyrmion moduli space into a 
real vector bundle with  natural $SO(2n+1)$ connection.
\end{abstract}
\bigskip


\end{titlepage}

It was already noted  \cite{Manton-2} that the solutions of
the Skyrme model \cite{Skyrme}, 
especially at the low energy, look like monopoles with the baryon number being
identified with the monopole topological number. This correspondence looks more clear
by using the moduli space approach   \cite{Manton}. 

In the paper by N.~Manton and B~Schroers  \cite{Manton-Schroers} the effect of
fermion interaction with BPS monopoles have been studied. The result is that the
$n$-monopoles moduli space turns into a vector bundle with $O(n)$
natural connection  
constructed from the fermion zero modes. 

In this note we would like to investigate if one could expand
the analogy between the moduli spaces of
the BPS monopoles and Skyrme model to the case when the interaction 
with the fermions is included.
The fundamental field $U({\bf x},t)$ of Skyrme's model is a map from coordinate 
space $M_4$ to the
configuration space of the mesons, in the simplest case given by the group $SU(2)$. 
The rescaled Lagrangian of the model (assuming zero bare mass) 
is \cite{Skyrme}
\begin{equation}                                    \label{Lagrang}
{\cal L} = \frac{1}{4}~ {\rm tr}~ \left( \partial_{\mu}U \partial^{\mu} U^{\dagger}\right) + 
\frac{1}{32}~ {\rm tr}~ \left[ (\partial_{\mu}U)  U^{\dagger}, (\partial_{\nu} U)  U^{\dagger}
\right]^2. 
\end{equation}
The first term here corresponds to the non-linear $\sigma$-model and the second 
one stabilizes the soliton solutions of the model.
In general it is useful to parametrise $U$ via 
a quartet of scalar fields  $(\phi_0, \phi^a)$, $ a=1,2,3$, as 
$$
U({{\bf r}}) = \left(\phi_0({\bf r}) + i \phi^a({\bf r}) \cdot
\sigma^a \right)
$$
with the constraint $\phi_0^2 + ({\phi^a})^2 = 1$. 

In this note we consider
the hedgehog configuration \cite{Skyrme} which in the static case is 
given by the ansatz
\begin{equation}                                    \label{hedgehog}
U({\bf r})_{_{n=1}} = e^{i (\sigma^a\cdot {\hat {r}^a}) F(r)} = 
\cos F(r) + i (\sigma^a \cdot {\hat {r}^a}) \sin F(r) .
\end{equation}
That defines the chiral angle $F(r)$. 
Here $\sigma^a$ are standard Pauli spin matrices and
the boundary conditions $F(0) = \pi, F(\infty) = 0$ correspond to the 
sector with  baryon (topological) number $n=1$. 

Note that the r.h.s. of (\ref{hedgehog}) appears as a special property 
of the $SU(2)$ group; generally speaking it is impossible to decompose 
a field $U$, which takes values in a Lie group $SU(N)$ with $N >2$, into 
a sum of $\sin$- and $\cos$-components.       

Extending the original Skyrme model to include the interaction with
fermions, the minimal chiral coupling of the fermion $SU(2)$ doublet
with Skyrme field is given by the Lagrangian
\begin{equation}             \label{skyrme-fermi}
{\cal L}_{int} =  
{\bar \psi} \left( i\gamma^{\mu} \partial_{\mu} + g U^{\gamma_5}\right)\psi =
{\bar \psi} \left[ i\gamma^{\mu} \partial_{\mu} + g \left( \phi_0 +
i\gamma_5 (\phi^a \cdot \sigma^a\right) \right]
\psi
\end{equation}
\noindent where 
$$ 
U^{\gamma_5} = \displaystyle \exp \{ 
i\gamma_5 \left(\sigma^a \cdot {\hat {
r}^a}\right) F(r)\} =  \displaystyle \frac{1+\gamma_5}{2} U +  \frac{1-\gamma_5}{2}
U^{\dagger}.
$$

The expression (\ref{skyrme-fermi}) can be considered as an analogue of the
Lagrangian of the monopole-fermion interaction. 
The obvious difference is the effect of
the $SU(2)$ magnetic monopole gauge potential $A_i$ which appears in the
covariant derivative $D_i = \partial_i + A_i$. 
Note that the topology of configurations 
is fixed by the
hedgehog ansatz in both cases and, on the other hand,
after a chiral rotation of the spinor field 
$$
\psi \to S \psi = \exp \left( -\frac{i}{2} \gamma_5~(\sigma^a
\cdot {\hat {r}^a}) F(r)\right) \psi
$$
\noindent the Lagrangian  (\ref{skyrme-fermi})
transforms to the form  
\begin{eqnarray}                  \label{Lagrang-psi-prime}
{\cal L}_{int}^{\prime} &=   {\bar \psi} 
\left[ i\gamma^{\mu}{\cal D}_{\mu} + g\right]
\psi  = {\bar \psi} \left[ i\gamma^{\mu}
\left( \partial_{\mu} + g S^{-1}\partial_{\mu} S  \right) + g\right] \psi 
\nonumber\\
 &={\bar \psi} \left[ i\gamma^{\mu}
\left(\partial_{\mu} + V_{\mu} + A_{\mu}\gamma_5\right) + g  \right] \psi,
\end{eqnarray}
where 
\begin{equation}
\begin{split}
V_{\mu} &= \frac{1}{2} \left(\xi \partial_{\mu} \xi^{\dagger} + \xi^{\dagger}
\partial_{\mu}\xi\right),\qquad 
A_{\mu} = \frac{1}{2} \left(\xi \partial_{\mu} 
\xi^{\dagger} - \xi^{\dagger}\partial_{\mu}\xi\right)\nonumber\\ 
\xi &= \exp\left\{
\frac{i}{2} (\sigma^a\cdot {\hat {r}^a}) F(r)\right\} = U^{1/2}.
\end{split}
\end{equation}
The effective chiral Lagrangian of that form was suggested in 
\cite{Diak-Eid}. Thus, the 
induced connection $S^{-1} \partial_\mu S$ generates an effective (pseudo)vector
potential of the Skirme model which relates the sectors with different baryon numbers.

Another difference between the cases of monopole-fermion and skyrmion-fermion 
interaction is the mass term. Indeed, using the Dirac representation of the $\gamma$
matrices, the Hamiltonian of the Dirac field chirally 
coupled to skyrmion can be written as \cite{Hiller,Bhad,Diakonov-2}:
{\small{
\begin{eqnarray}                                    \label{ferm-skirm}
&H \psi = \left( \alpha^a \cdot {p}^a + g\beta \cos F(r)
+ ig \gamma_5~({\tau^a
\cdot {\hat r}^a}) \sin F(r) \right) \psi 
\equiv \\[3pt]  
&\left( 
\begin{array}{cc} 
M(r)   & {\cal D}  \\[3pt] 
 {\cal D}^{\dagger} &  -M(r)
\end{array} \right) \psi 
\equiv 
\left( 
\begin{array}{cc} 
g \cos F(r)   & \sigma^a \cdot {p}^a + {ig}(\tau^a
 \cdot {\hat {r}^a}) \sin F(r)  \\[3pt] 
 \sigma^a \cdot {p}^a - {ig}(\tau^a
 \cdot {\hat {r}^a} ) \sin F(r) & -g \cos F(r)
\end{array} \right) \psi = E \psi \nonumber
\end{eqnarray}
}}
where ${\bf p} = -i\nabla$ and
the Dirac operator is defined as 
$$
{\cal D} =  \sigma^a \cdot {p}^a + {ig}(\tau^a
\cdot {\hat {r}^a}) \sin F(r) . 
$$

Thus, the equation (\ref{ferm-skirm}) describes the spinor field which
has a space 
dependent dynamical complex mass.  Its counterpart is 
the Hamiltonian of the monopole-fermion interaction
{\small{
\begin{eqnarray}                                    \label{ferm-mono}
&H_{BPS} \psi = \left( \alpha^a \cdot \left({p}^a + {\vec A}\right)  
+  iq\gamma_5~\beta (\tau^a
 \cdot {\hat {r}^a}) H(r) \right) \psi 
\equiv \\[3pt]  
&\left( 
\begin{array}{cc} 
0   & {\cal D'}  \\[3pt] 
 {\cal D'}^{\dagger} &  0
\end{array} \right) \psi 
\equiv 
\left( 
\begin{array}{cc} 
0   & \sigma^a \cdot ({p}^a + {A}^a) + {iq}(\tau^a
\cdot {\hat {r}^a})H(r)  \\[3pt] 
 \sigma^a \cdot ({p}^a + {A}^a) - {iq}(\tau^a
 \cdot {\hat {r}^a} )H(r) & 0
\end{array} \right) \psi = E \psi \nonumber
\end{eqnarray}
}}
where we suppose that the fermion mass is entirely  
due to the pseudoscalar 
coupling between the Higgs field and fermions. Here
the Dirac operator is 
$$
{\cal D'}=\sigma^a \cdot ({p}^a + {A}^a) +
iq \displaystyle (\tau^a \cdot {\hat {r}^a})H(r)$$ 
and 
$$
H(r) = \left(\frac{1}{r} - \cosh r\right), \quad {\rm  
and} \quad A_i^a = \varepsilon_{iab}\frac{r^b}{r}\left(\frac{1}{r} - 
\frac{1}{\sinh r}\right)
$$ 
are well known BPS monopole solutions.

Obviously, there are common features of the monopole-fermion and skyrmion-fermion interaction.
In both cases the Hamiltonians are commute with the generalized angular 
momentum ${\bf J} = {\bf L} + {\bf T} + {\bf S}$ which composed operators of 
standard orbital momentum ${\bf L}$, isospin ${\bf T}$ and spin ${\bf S}$. 
However, the Hamiltonian $H$ of equation 
(\ref{ferm-skirm}) also commutes with the Dirac parity operator defined by
$$
\psi({\vec r}) \to \beta\psi(-{\vec r})
$$ 
while the Hamiltonian of the fermion-monopole interaction is
invariant under a joint parity transformation and magnetic charge conjugation.

The most important, from point of view of our consideration, are 
the low-energy modes of the Hamiltonian (\ref{ferm-skirm}). 
One could expect that
such a low energy state has spherical symmetry in the sense that it is invariant
under a combined spatial and isospin rotation.  Then
the Dirac $j=0$ positive parity spinor can be written as
\begin{equation}                               \label{decouple}
\psi ({\bf r}) = \frac{1}{r}\left( 
\begin{array}{c} i \rho(r)  \\
(\tau^a \cdot {\hat {r}^a})\lambda(r)\end{array} \right)|\chi>
\end{equation}
where $|\chi>$ is the spin-isospin spinor, i.e. $\left(\tau^a + \sigma^a
\right) |\chi > = 0$ that means $|\chi > = -i \sigma_2 \otimes I_2$.
  
The equations for spinor components of the $\psi$ are straightforward:
\begin{eqnarray}
\left(\frac{d}{dr} + g \cos F - \frac{1}{r}\right) \rho =& \left(E - g\sin
F\right) \lambda;
\nonumber\\  
\left(\frac{d}{dr} - g \cos F + \frac{1}{r}\right) \lambda =& -\left(E + g\sin
F\right) \rho
\end{eqnarray}
Note, that the effect of the mass term is that, even in the case $E=0$, 
these equations do not decouple into two
independent equations for the spinors $\rho$ and $\lambda$ as it was in 
the case of monopole-fermion interaction.
It means, that a chirally coupled skyrmion may or may not give rise to a zero energy
fermion mode depending on the strength of the coupling (effective mass)  
\cite{Wil-MacKen}. The investigation of the constraints under which a zero energy normalisable
solution of the system (\ref{decouple}) exists have been done in
\cite{Hiller}-\cite{Carena}. 

In contrast with the BPS monopole-fermion system there is no analytical solution of this
equation and in order to simplify the numerical calculations some ansatzs for the shape function
(the chiral angle $F(r)$) were implemented. 
The general result of such calculations is that there is a
spectral flow of the Hamiltonian (\ref{ferm-skirm}); 
for some small value of parameter $g$ a single
eigenvalue of $H$ emerges from the positive energy continuum, crosses zero and goes to the negative
continuum if $g$ increases \cite{Niemi}. In contrast, in the case of a BPS monopole, the
Hamiltonian (\ref{ferm-mono}) has a single zero energy eigenvalue that does not depend from the
coupling constant \cite{Manton-Schroers}. Nevertheless the index theorem also can be applied to the
Hamiltonian of the form  (\ref{ferm-skirm}) \cite{Niemi}: Writing it in the form
\begin{equation}
H = \left( 
\begin{array}{cc} 
M   & {\cal D}  \\[3pt] 
 {\cal D}^{\dagger} &  -M
\end{array} \right) = H_0 + M\sigma_3, \qquad H^2 \ge M^2,
\end{equation}
we see that ${\rm Ind} [H_0] = {\rm Dim~Ker} [{\cal D}^{\dagger}] -  {\rm Dim~Ker}[{\cal D}] 
= n$.
The difference from the BPS monopole - fermion system is that zero mode of the Dirac operator
does not lies in the the kernel of ${\cal D}$.

In order to describe the moduli space of the Skyrme model note that
its symmetry group is $G = E_3 \times SO(3)$, where 
Euclidean group $E_3 = SO(3) \times R_3$. This group is  9-dimensional but, when 
it acts on a field of
the hedgehog form  (\ref{hedgehog}), the spatial and the isospin rotations are equivalent. 
Hence the orbit of the
standard skyrmion under symmetry group is a six-dimensional manifold ${\cal M}_1$.
This is the manifold which is used as the moduli space in the $n=1$ topological
sector. It is diffeomorphic to $R_3 \times SO(3)$ and its elements are fully specified
by their position and orientation. On the moduli space  ${\cal M}_1$ the potential
energy is constant and equal to the skyrmion's rest mass $M$. The induced dynamics is
determined by the kinetic energy which can be found \cite{Adkins} by inserting the
adiabatically rotating and moving ansatz into (\ref{Lagrang}):
\begin{equation}                     \label{rotate}
U({\bf r},t) = S(t)U_0({\bf X(t)} - {\bf r})S(t)^{\dagger}, \quad \psi({\bf r},t) \to
 S(t) \psi ({\bf r},t),
\end{equation}
where $S(t)$ is an $SU(2)$ cranking matrix and ${\bf X}(t)$ 
are collective coordinates on $R_3$. 
Exploiting the identity 
$$ 
S^{\dagger} {\dot S} = \frac{\tau^a}{2} ~{\rm tr} \left(\tau^a S^{\dagger} 
{\dot S} \right)
$$ 
after a lengthy calculation one obtains the induced
Lagrangian 
\begin{equation}               \label{eff-lang}
L =  \frac{1}{2} M{\dot{\bf X}}^2 +   
\Lambda ~{\rm tr}~\left({\dot S}^{\dagger} {\dot S} \right) +
\frac{i}{2} ~{\rm tr} \left(\tau^a S^{\dagger} {\dot S} \right) {\Sigma}^a,
\end{equation}
where  
$               
\Sigma^a = \int d^3x~ \psi^{\dagger} \tau^a \psi$ is the fermions contribution
to the total angular momentum of the configuration,
$\Lambda$ is the moment of inertia associated with the collective rotations,
and the fermion kinetic energy term is ignored as well as the skyrmion rest 
mass $M$. 

Let us remind that the spatial rotations act on
$S(t)$ by right multiplication with $SU(2)$ matrix, 
while isospin rotations act by 
left multiplication. According to the Noether's theorem, the invariance
of the effective Lagrangian with respect to the 
space rotations $S(t) \to S(t) 
\exp\{i\omega^a \sigma^a/2\}$ leads to the conservation of the total 
momentum 
\begin{equation}               \label{spin}
{\bf J} = i\Lambda  ~{\rm tr}~\left({\pmb\sigma} {S}^{\dagger} {\dot S} \right) - 
{\pmb\Sigma},
\end{equation}
and the effective Lagrangian (\ref{eff-lang}) can be written as
\begin{equation}                  \label{Lagr-mod}  
L = \frac{1}{2} M{\dot{\bf X}}^2 + \frac{1}{2\Lambda}\left[ 
{\bf J}^2 - {\pmb\Sigma}^2\right]
\end{equation}

Let us analyse the structure of that Lagrangian. 
The common parameterization of the $SU(2)$ cranking matrix $S(t)$ is 
\begin{equation}
S(t) = \exp\{\frac {i}{2} \sigma_k \omega_k(t)\} = a_0(t) + i  \sigma_k a_k (t) =
 \left( 
\begin{array}{cc} 
a_0 + ia_3  & a_2 + ia_1  \\[3pt]
-a_2 + ia_1 &  a_0 - ia_3
\end{array} \right)
\end{equation}
where $a_{\mu}^2 =1$ that is $a_\mu$ are the coordinates of a point on a sphere $S_3$. 
An alternative is to 
introduce the Euler angles $\alpha, \beta, \gamma$ on the
three-sphere according to:
\begin{equation}
S = \left( 
\begin{array}{cc} \cos \frac{\beta}{2}e^{\frac{i}{2}(\gamma + \alpha)} & \sin 
\frac{\beta}{2}e^{\frac{i}{2}(\gamma - \alpha)}
\\[3pt] 
- \sin \frac{\beta}{2}e^{-\frac{i}{2}(\gamma - \alpha)} &\cos \frac{\beta}{2}
e^{-\frac{i}{2}(\gamma + \alpha)} 
\end{array} \right)
\end{equation}
Recall that left and right rotations on $SO(3)$ in terms of the  Euler angles are 
generated by one-forms which are familiar in the analysis of rigid body rotations:
$$
L =   d S \cdot S^\dagger = \frac{i}{2} \sigma_k L_k;\qquad
R = S^\dagger \cdot d S = \frac{i}{2} \sigma_k R_k
$$  
with the property $S^\dagger \cdot d S = - dS^\dagger \cdot S$. Here the components 
of the velocities are 
\begin{eqnarray}
L_1 = {\dot \beta} \sin \alpha - {\dot \gamma} \sin \beta \cos \alpha; &\quad& 
R_1 = - {\dot \beta} \sin \gamma + {\dot \alpha} \cos \gamma \sin \beta;\nonumber\\
L_2 = {\dot \beta} \cos \alpha + {\dot \gamma} \sin \beta \sin \alpha; &\quad&
R_2 = {\dot \beta} \cos \gamma + {\dot \alpha} \sin \gamma \sin \beta;\nonumber\\
L_3 = {\dot \alpha} + {\dot \gamma} \cos \beta; &\quad& 
R_3 = {\dot \gamma} + {\dot \alpha} \cos \beta
\end{eqnarray}
Thus the straightforward calculation yields two equivalent forms of the Lagrangian of
collective motion on one-skyrmion moduli space written via the operators of left and 
right rotation  respectively: 
\begin{eqnarray}    \label{L}
L &=& \frac{1}{2} M{\dot{\bf X}}^2 +  
\frac{\Lambda}{2} L_k^2 + \frac{1}{2}{\Sigma}^k L_k; \nonumber\\
L &=& \frac{1}{2} M{\dot{\bf X}}^2 +  
\frac{\Lambda}{2} R_k^2 - \frac{1}{2}{\Sigma}^k R_k
\end{eqnarray}
Furtermore, if we consider only the contribution of the low-energy fermionic 
quasi-zero  modes  (\ref{decouple}) and impose the normalization condition 
$\int dr (\rho^2 + \lambda^2) = 1$, the fermionic contribution to the 
angular momentum becomes $\pmb \Sigma = \pmb \sigma$. Thereafter we make 
no difference between the spin and isospin matrices.

Note that in the parameterization by the Euler angles 
the metric on the group  manifold is non-diagonal
\begin{equation}
dS^2 = \frac{1}{4} d\alpha^2 +  \frac{1}{4} d \beta^2 +  \frac{1}{4} d \gamma^2 +  
\frac{1}{2} \cos\beta~d\alpha d\gamma
\end{equation}
Therefore it would be more convenient to introduce the orthogonal coordinates
\begin{equation}
\psi = \gamma + \alpha; \qquad \chi = \gamma - \alpha
\end{equation}
in terms of which the metric becomes
\begin{equation}
dS^2 = \frac{1}{4} d \beta^2 + 
\frac{1}{4} d \psi^2 \cos^2 \frac{\beta}{2} + \frac{1}{4} 
d \chi^2 \sin^2 \frac{\beta}{2}
\end{equation}

Consider now the complete path around the group manifold. Obviously it is parameterized 
by  the values of the Euler angles ranging within intervals $\beta \in [0,\pi];~ \alpha \in [0,2\pi];~ 
\gamma \in [0,2\pi]$. However, 
since these rotations act independently on the chiral components of spin-isospinor wave function 
$\psi_R \to U_R \psi_R;~ \psi_L \to U_L \psi_R$, the lower energy state transforms as
$$ 
U_R \mid \chi > = - U_L |\chi >
$$
Therefore,  
there is a non-trivial holonomy on the sphere and we obtain
$SO(3)$ bundle over the moduli space of the 
Skyrme model with a unit topological charge. 
Indeed, the corresponding $SO(3)$ connection can be easily calculated if we consider the canonical
momenta which correspond to the effective Lagrangian (\ref{L}):
 \begin{eqnarray}
{\bf P} &=& 
M{\dot {\bf X}}; \qquad \Pi_\alpha = \Lambda({\dot \alpha} + {\dot \gamma} \cos \beta) + {\Sigma_3};\nonumber\\
\Pi_\beta &=& \Lambda {\dot \beta} + {\Sigma_1}\sin \alpha - \Sigma_2 \cos \alpha ;\nonumber\\
\Pi_\gamma &=& \Lambda ({\dot \gamma} + {\dot \alpha}) - \Sigma_1 \sin \beta \cos \alpha - \Sigma_2 \sin \beta \sin \alpha
+ \Sigma_3 \cos \beta \nonumber
\end{eqnarray}
and one can see that the  fermion interaction term gives rise to the 
effective non Abelian gauge potential $ A_i = {A_{i}^a}\Sigma^a$
\begin{eqnarray}              \label{Berry}
A_\alpha = \frac{1}{2} A_\alpha^a\sigma^a = \frac{1}{2} \left( 
\begin{array}{cc} 
1   & 0  \\[3pt] 
0  &  -1
\end{array} \right);&&\quad A_\beta = \frac{1}{2}A_\beta^a\sigma^a =
\frac{i}{2}\left( 
\begin{array}{cc} 
0   & e^{-i\alpha}  \\[3pt] 
- e^{i\alpha} &  0
\end{array} \right);\nonumber\\
A_\gamma &=&  \frac{1}{2}A_\gamma^a\sigma^a = - \frac{1}{2}\left( 
\begin{array}{cc} 
-\cos \beta   & \sin \beta e^{-i\alpha}  \\[3pt] 
\sin \beta e^{i\alpha} &  \cos \beta
\end{array} \right)
\end{eqnarray}
The mechanism of the effective potential generation is well known \cite{Aitch},
\cite{Moody}. It is connected with the nontrivial holonomy on sphere $S_3$. Indeed,
the well-known Hopf fibration arises if one considers the $S_3$ as a principal fibre bundle with
base $S_2$ and a structure group $U(1)$. Then 
the additional term, which appears due to the fermions in the 
formula (\ref{eff-lang}), on the sphere $S_2$  can be identified as 
the Balachandran-Aitchison monopole effective Lagrangian \cite{Aitch,Bal}:
\begin{equation}            \label{L-berry-3}
L_{eff} = \frac{i}{2} s~{\rm tr} \left(\sigma_3 S^{\dagger} {\dot S} \right).
\end{equation}
This term is in fact a Wess-Zumino-type one. Indeed, as it was suggested by Diakonov
and Petrov \cite{Diakonov-3}, one can introduce a unit three-vector 
$n^a = \displaystyle \frac{1}{2}~{\rm tr} \left(S \sigma^a S^{\dagger}
\sigma_3\right)$ and rewrite the effective Lagrangian (\ref{L-berry-3}) as
\begin{equation}
 L_{eff} = i \frac{s}{4}~\int d\sigma ~\varepsilon_{abc}\varepsilon_{ij}~ n^a
\partial_i n^b \partial_j n^c,
\end{equation}
\noindent where the surface integral actually is a full derivative and define the
topological charge on the group space. 
Then the charge quantization condition
takes place. 

Indeed, it is clear that the model is still invariant under $U(1)$ time-dependent 
gauge transformation $U \to U\exp\{i{\hat Q} \alpha(t)\} = U \exp\{i \sigma_3 \alpha (t)\}$ where 
${\hat Q} =2s$ is $U(1)$ generator. 
The Lagrangian (\ref{L-berry-3})
transforms as $L_{eff} \to L_{eff} - 2s {\dot \alpha}$ that is full time derivative.
Thus the symmetry group is  $SU(2)/U(1)$ rather then $SU(2)$ and in quantum theory 
the physical states $\Phi$ that are eigenfunctions of the Hamiltonian on moduli space have to be restricted to
be also eigenfunctions of operator ${\hat Q}$:   
\begin{equation}           \label{Gauss}
{\hat Q} \Phi = 2s \Phi.
\end{equation}
 As a result the parameter $s$ have 
to be quantized as $s = n/2, n \in Z$ where $n$ is a winding number
associated with above mentioned $U(1)$ gauge degrees of freedom 
\cite{Aitch}. 
That means that coupling
of the skyrmion with fermions could affect the statistics. Indeed, under $2\pi$ rotations 
$\Phi \to \Phi \exp \{i\pi \sigma_3\} = \Phi \exp \{i\pi {\hat Q}\}$.  The condition (\ref{Gauss}) means that
$\Phi \to \Phi \exp \{2i\pi s\}$, i.e skyrmion coupled with odd number of quarks transforms as a fermion and
as a boson if it is coupled with even number of quarks.    

In order to generalize this construction to the case of the 
skyrmion with topological charge
$n$ let us consider quantum mechanics on the moduli space. 
The canonical quantization prescription gives the quantum Hamiltonian 
\begin{equation}         \label{Ham}
H = -\frac{1}{2M}\frac{\partial^2}{\partial {\bf X}^2} - 
\frac{1}{2 \Lambda}\left(
{\bf J}^2 - (\pmb\Sigma)^2\right) \equiv \nabla_{\alpha}\nabla^{\alpha}
\end{equation} 
\noindent where ${\bf J}$ is the angular momentum operator defined by 
eq.(\ref{spin}) and 
we introduce the shorthand $\nabla_{\alpha}$, $\alpha = 1,2\dots 6$, 
for the
covariant derivative associated with the connection on the moduli 
space of Skyrme model
${\cal M}$ parameterized by the set of coordinates 
$\xi_{\alpha} =  ({X}_k, \omega_k)$.

Now one can apply the adiabatic approximation for the eigenfunctions of the 
Hamiltonian  (\ref{Ham}) taking into account the fermionic degrees of freedom: 
\begin{equation}
\Phi (\xi_{\alpha}, {\bf r}) = \Psi(\xi_{\alpha}) \psi(\xi_{\alpha}, {\bf r})
\end{equation}
where $\psi(\xi_{\alpha}, {\bf r})$ is the single fermion quasi-zero 
mode (\ref{decouple}). 

Using the Born-Oppenheimer
adiabatic approximation \cite{Manton-Schroers},\cite{Moody} we can consider 
these 
fermionic degrees of freedom as the ``fast'' variables and 
the ``slow'' variables, which 
are the coordinates $\xi_{\alpha}$ 
on the 6-dimensional moduli space ${\cal M}_1$, describe the effective quantum dynamic of
the skyrmion. Multiply the Schr\"odinger equation with the 
Hamiltonian (\ref{Ham}) 
$$H \Phi = \nabla_{\alpha}\nabla^{\alpha}\Phi = E\Phi
$$
on the left by $\psi^{\dagger}(\xi_{\alpha}, {\bf r})$ and integrate over fermionic
coordinates ${\bf r}$ one can obtain exploiting the orthogonality of 
$\psi (\xi_{\alpha}, {\bf r})$:
\begin{equation}
\nabla_{\alpha}\nabla^{\alpha}\Psi + 2 < \psi, \nabla_{\alpha}\psi> 
\nabla^{\alpha}\Psi + \Psi <\psi, \nabla_{\alpha}\nabla^{\alpha}\psi>
= E \Psi.
\end{equation} 
Introduce a local gauge potential ${\cal A}^{eff}_{\alpha} = 
i \langle \psi,\nabla_{\alpha} \psi \rangle$ and neglecting the transitions between
the fermions levels, that is the Born-Oppenheimer approximation, we can see that    
the matrix-valued Hamiltonian sandwiched between the "fast" degrees of freedom 
becomes 
\begin{equation}
H_{eff} =  \left( \nabla_{\alpha} - i{\cal A}^{eff}_{\alpha} \right)^2
\end{equation}
This result can be obtained also if we note that adiabatically rotating 
fermionic field  of the form (\ref{rotate}) can be expanded  
in the complete set of
unrotating zero modes $\psi_{n}({\bf x})$. According the index theorem the 
number of these modes is equal to the topological charge $n$ and
therefore the Hamiltonian $H$ for each band $n$
may be regarded as an element 
of the algebra $SO(2n+1)$.

\medskip\noindent{\bf Acknowledgements}.

This research is inspired by numerous 
discussions with Steffen Krusch \cite{Kirsh} and  
with Nick Manton. Part of this work was done while the author was at DAMTP,  
University of Cambridge. I would like to acknowledge the hospitality at the Abdus Salam 
International Center for Theoretical Physics where this work was completed.

\end{document}